\documentclass[12pt]{iopart}
\usepackage{epsfig}
\usepackage{iopams}
\bibstyle{unsrt}
\newcommand{\be}{\begin{equation}}
\newcommand{\ee}{\end{equation}}
\newcommand{\bea}{\begin{eqnarray}}
\newcommand{\eea}{\end{eqnarray}}

\newcommand{\dd}{\rmd}
\newcommand{\dyad}[1]{\buildrel{\leftrightarrow}\over{\mathbf{#1}}}
\newcommand{\mb}[1]{\bi{#1}}
\newcommand{\rot}{\mathrm{Curl}}
\newcommand{\im}[1]{\Im\mathrm{m}\left\{ #1 \right\}}
\newcommand{\intz}{\int_0^\infty \dd \zeta}
\newcommand{\uv}[1]{\hat{\mb{#1}}}
\newcommand{\vkl}{\mb{k}_\perp}
\newcommand{\rmh}{\rm{h}}
\newcommand{\kl}{k_\perp}
\newcommand{\sEM}{{\sum_{q=\mathrm{TE}}^\mathrm{TM}}}
\newcommand{\kj}{\kappa_j}
\newcommand{\ki}{\kappa_1}
\newcommand{\kii}{\kappa_2}
\newcommand{\kg}{\kappa_g}
\newcommand{\Diq}{\Delta_{1q}}
\newcommand{\Diiq}{\Delta_{2q}}

\begin{document}
\title[Casimir attraction in plane parallel systems]{Casimir attraction in multilayered plane parallel magnetodielectric systems}
\author{Simen A Ellingsen\footnote{Permanent address: Department of War Studies, King's College London, Strand, London WC2R 2LS, United Kingdom}}
\address{Department of Energy and Process
Engineering}
\address{Norwegian University of Science and
Technology, N-7491, Trondheim, Norway}
\ead{\emph{simen.ellingsen@kcl.ac.uk}}

\date\today

\begin{abstract}
A powerful procedure is presented for calculating the Casimir attraction between plane parallel multilayers made up of homogeneous regions with arbitrary magnetic and dielectric properties by use of the Minkowski energy-momentum tensor. The theory is applied to numerous geometries and shown to reproduce a number of results obtained by other authors. Although the various pieces of theory drawn upon are well known, the relative ease with which the Casimir force density in even complex planar structures may be calculated, appears not to be widely appreciated, and no single paper to the author's knowledge renders explicitly the procedure demonstrated herein. Results may be seen as an important building block in the settling of issues of fundamental interest, such as the long-standing dispute over the thermal behaviour of the Casimir force or the question of what is the correct stress tensor to apply, a discussion re-quickened by the newly suggested alternative theory due to Raabe and Welsch.

\end{abstract}

\maketitle

\section{Introduction}
\label{sec:intro}

Over the last decade or so enormous progress has been made in experimental tests of the Casimir effect \cite{Casimir:1948}. This macroscopic manifestation of quantum electrodynamics, once something of a curiosity subject mainly to the scrutiny of a few theorists, has been measured with high precision and is now spoken of as possibly exploitable in nanoelectromechanical applications \cite{Chan:2001}. 

Although only a single experiment has so far employed parallel plates and with moderate accuracy \cite{Bressi:2002}, experiments employing other geometries (typically a sphere and a plate) have normally had to resort to planar geometries for theoretical support, accompanied by the neccessary corrections to account for curved surfaces (see e.g.\ \cite{Lamoreaux:2004}). 

The individual pieces of theory assembled in this paper are not in themselves new; the paper draws heavily on several references, many of which more than a decade old. The theory of Green's functions in a dielectric multilayer was treated by Toma\v{s} eleven years ago \cite{Tomas:1995}, building in turn on previous work by Mills and Maradudin \cite{Mills:1975} two decades earlier. Companioned by the now classical theory by Lifshitz and co-workers \cite{Lifshitz:1961} and standard optical theory of reflection it provides all the necessary tools. Desipte this fact however, the ease with which the Casimir forces in plane parallel systems may be expessed appears not to be commonly recognised, although it has been implicitly employed by Toma\v{s} (\cite{Tomas:2002} and later papers). Furthermore, no publication exists to the author's knowledge, explaining explicitly the procedure derived and demonstated herein.

This paper provides background theory to aid the settlement of at least two ongoing disputes in the Casimir branch. Firstly, the as yet unsettled disagreement over the temperature effect of the Casimir force (for a recent discussion see \cite{Brevik:2006} and references therein); numerical and theoretical treatment of the expressions obtained e.g.\ in chapter \ref{ch_coefficients} using dispersion data for real materials provide predictions to settle experimentally the existence or non-existence of the large thermal variations of the force upheld by many. Secondly, doubts have been raised recently about the applicability of the Minkowski stress tensor and an alternative, Lorentz force-based tensor was suggested \cite{Raabe:2005}, in turn disputed by Pitaevskii \cite{Pitaevskii:2006}. A procedure similar to that presented here seems to have been employed by Toma\v{s} in his calculations of the effects of the Lorentz-type tensors \cite{Tomas:5z, Tomas:4z}. The discussion of the applicability or indeed correctness of their theory, however, is not within the scope of this paper.

We have structured the paper as follows. In chapters \ref{ch_background} to \ref{ch_coefficients} the background theory of Green's function calculation of the Casimir attraction is derived briefly, arriving at equation (\ref{eq_F}), the main result of the paper. In chapter \ref{ch_configs} we demonstrate the strength of the procedure by using it to readily reproduce an array of previous results in various configurations: two half-spaces, a plate and a wall, a plate in a cavity and two plates.

Many detailed calculations which are straightforward in principle have been omitted. For details, the reader may refer to \cite{Ellingsen:master}.

\section{Background theory: force on an interface}\label{ch_background}

When electro- and magnetostrictive contributions are neglected, forces acting inside magnetodielectric media assuming no net external charge or currents are present, may in general be expressed through space components of the Minkowski energy momentum tensor \cite{Brevik:1979}. Assuming isotropic, homogeneous and linear media the electromagnetic force density acting at position $\mathbf{r}$ is
\be\label{eq_f}
  f_i(\mb{r}) = \partial_k T_{ik}= -\frac{1}{2}\epsilon_0 \mb{E}^2 \partial_i \epsilon(\mb{r}) -\frac{1}{2}\mu_0 \mb{H}^2 \partial_i \mu(\mb{r}),
\ee
where $T_{ik}$ is the Maxwell stress tensor,
\be
  T_{ik}=E_iD_k + H_iB_k - \frac{1}{2}\delta_{ik}(\mb{E}\cdot\mb{D}+\mb{H}\cdot\mb{B}),
\ee
where indices $i,k \in \{x,y,z\}$ denote Cartesian vector components. We have suppressed the frequency dependence of the permittivity and permeability, respectively $\epsilon$ and $\mu$, both defined relative to vacuum so that $D_i = \epsilon_0 \epsilon E_i$ and $B_i = \mu_0\mu H_i$.

We introduce the classical Green's dyadic $\Gamma_{ik}$ according to the convention of Schwinger and co-workers \cite{Schwinger:1978} defined according to
\be\label{def_Gamma}
  \mb{E}(x) = \frac{1}{\epsilon_0}\int \dd^4x' \dyad{\Gamma}(x,x')\cdot \mb{P}(x')
\ee
where $x=(\mb{r},t)$. Due to causality, $t'$ is only integrated over the region $t'\leq t$: the polarisation at a time $t'$ cannot influence the resulting electric field at time $t$ prior to $t'$. The definition (\ref{def_Gamma}) ensures that $\Gamma$ is a generalised susceptibility. It is well known that according to Maxwell's equations, $\Gamma$ satisfies
\be\label{eq_basic_Gamma}
  \nabla\times\nabla\times \dyad{\Gamma}(\mb{r},\mb{r}'; \omega) - \frac{\epsilon(\mb{r})\mu(\mb{r})\omega^2}{c^2}\dyad{\Gamma}(\mb{r},\mb{r}'; \omega) = \frac{\mu(\mb{r})\omega^2}{c^2}\delta(\mb{r}-\mb{r}')\dyad{1},
\ee
where we have performed a Fourier transformation according to
\be \label{eq_Fourier_w}
  \dyad{\Gamma}(x,x') = \int_{-\infty}^\infty \frac{\dd \omega}{2\pi}\rme^{-i\omega \tau}\dyad{\Gamma}(\mb{r}, \mb{r}';\omega),
\ee
with $\tau \equiv t-t'$.

Invoking the fluctuation-dissipation theorem in a standard manner (e.g.\ \cite{Lifshitz:1961}), yields\footnote{Compared to ref.\ \cite{Lifshitz:1961}, $\Gamma=-\frac{\omega^2}{\hbar c^2}D$.}
\numparts
\begin{eqnarray}
  \fl\rmi\langle E_i(\mb{r})E_k(\mb{r}')\rangle_\omega &= \frac{\hbar}{\epsilon_0}\coth\left(\frac{\hbar\omega}{2k_BT}\right)\im{\Gamma_{ik}(\mb{r}, \mb{r}'; \omega)}\label{eq_EE}\\
  \fl\rmi\langle H_i(\mb{r})H_k(\mb{r}')\rangle_\omega &= \frac{\hbar}{\mu_0}\coth\left(\frac{\hbar\omega}{2k_BT}\right)\frac{c^2}{\mu\mu'\omega^2}\rot_{ij}\rot_{kl}'\im{\Gamma_{jl}(\mb{r}, \mb{r}'; \omega)},\label{eq_HH}
\end{eqnarray}
\endnumparts
where we have used the notation $\rot_{ik}\equiv \epsilon_{ijk}\partial_j$ ($\epsilon_{ijk}$ being the Levi-Civita symbol and summation over identical indices is implied), $\rot_{ik}' \equiv \epsilon_{ijk}\partial_j'$ where $\partial_j'$ is differentiation with respect to component $j$ of $\mb{r}'$, and $\mu'\equiv \mu(\mb{r}')$. The brackets denote the mean value of the $\omega$ fourier component of the field component products with respect to fluctuations. For now let us assume $T=0$ for simplicity.

Following the Lifshitz procedure we introduce the useful quantities
\numparts
\begin{eqnarray}
  \Gamma^E_{ik}(\mb{r}, \mb{r}'; \omega) &\equiv& \Gamma_{ik}(\mb{r}, \mb{r}'; \omega);\\
  \Gamma_{ik}^H(\mb{r}, \mb{r}'; \omega) &\equiv& \frac{c^2}{\omega^2}\rot_{il}\rot_{km}'\Gamma_{lm}(\mb{r}, \mb{r}'; \omega),
\end{eqnarray}
\endnumparts
and replace the (fluctuation averaged) field components contained in $T_{ik}$ with their Green's function equivalents. 

Consider now a sharp interface between two different magnetodielectric media such as described above. The force density acting on the interface equals the net $zz$-component of $T_{ik}$ when $\mb{r}\to \mb{r}'$ at the surface, averaged with respect to fluctuations, that is the net flow of momentum through the interface when all unphysical contributions have been subtracted, i.e.\ all that flowing in equal amounts both ways across the interface. The resulting force expression is \cite{Lifshitz:1961}  (the original expression has been generalised to allow $\mu\neq 1$)
\be \label{eq_Tzz_Lifshitz}
  \fl\langle \mathcal{F}_z\rangle = \hbar \int_0^\infty \frac{\dd \zeta}{2\pi}\left[\epsilon(\Gamma_{xx}^{E,\rmh-}+\Gamma_{yy}^{E,\rmh-}-\Gamma_{zz}^{E,\rmh-})+\frac{1}{\mu}(\Gamma_{xx}^{H,\rmh-}+\Gamma_{yy}^{H,\rmh-}-\Gamma_{zz}^{H,\rmh-}) \right]_{\mb{r}=\mb{r}'},
\ee
where a standard frequency rotation to $\omega =i\zeta$ has been performed. The dependence of the components of $\Gamma$ on $\mb{r}, \mb{r}'$ and $\omega$ has been suppressed and superscript $\rmh$ indicates that only the homogeneous solution of (\ref{eq_basic_Gamma}) is included. A solution of (\ref{eq_basic_Gamma}) generally takes the form $f(z-z') + g(z+z')$ and the superscript `$-$' denotes that terms of $\Gamma$ dependent on $z+z'$ are discarded\footnote{The terms of $\Gamma$ omitted from (\ref{eq_Tzz_Lifshitz}) are non-physical in our formalism. The particular solution of (\ref{eq_basic_Gamma}) equals the Green's function in an infinitely large and homogeneous magnetodielectric, and is thus geometry independent. The $z+z'$-terms, argues Lifshitz, make no contribution to the net flux of momentum inside a single homogeneous medium, and it is straightforward to show formally that they make zero contribution to (\ref{eq_Tzz_Lifshitz}) if included \cite{Ellingsen:master}. These terms may be understood as a constant background of radiation contributing to what Milton in his book refers to as `bulk energy' \cite{Schwinger:1978}, generally of different value in different materials. Notably, if one followed the procedure Schwinger, Milton et.al.\ use to include the stress tensor, all terms of the homogeneous solution must be included since the Green's function is evaluated on both sides of an interface, i.e.\ in two different media. Both approaches yield the same end result as they should.}. The limit $\mb{r}\to\mb{r}'$ is taken so that (\ref{eq_Tzz_Lifshitz}) is evaluated entirely on one side of the interface. 

Prior to solving (\ref{eq_basic_Gamma}) explicitly in some planar geometry, we introduce one further Fourier transformation:
\be
  \dyad{\Gamma}(\mb{r},\mb{r}';\omega)= \int \frac{\dd^2 \kl}{(2\pi)^2}\rme^{i\vkl\cdot(\mb{r}_\perp-
  \mb{r}_\perp')}\dyad{\Gamma}(z,z';\vkl,\omega),
\ee
where the subscript $\perp$ denotes a direction perpendicular to the longitudinal or $z$-axis, i.e.\ in the $xy$-plane.

\section{Green's functions in a multilayer geometry}
\label{Sec:Green}

\begin{figure}
  \begin{center}
  \includegraphics[width=4in]{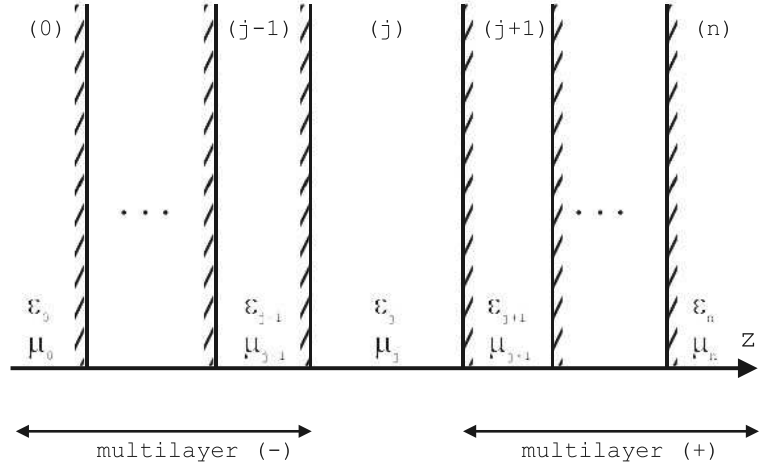}
  \caption{The multilayer geometry as discussed. We need to calculate the homogeneous part of the Green's function in layer $(j)$.}\label{fig_multilayer}
  \end{center}
\end{figure}

We regard a multilayered geometry with a total of $n+1$ layers such as depicted in figure \ref{fig_multilayer}. Each layer is assumed homogeneous, isotropic and of infinite transverse size. The thickness of some layer $(l)$ is denoted $a_l$. The Casimir attractive force per unit transverse area between multilayers $(+)$ and $(-)$ bordering on either side of some (not arbitrarily chosen) layer $(j)$ so that $0<j<n$ may be calculated using (\ref{eq_Tzz_Lifshitz}) by evaluating the Green's function near any of the boundaries of layer $(j)$. In a real setting, layers $(0)$ and $(n)$ will both typically be vacuum or air, and it should be obvious that the two multilayers exert forces on each other in a resiprocal manner (there is a subtlety when material $0$ does not equal material $n$ as discussed below).

Toma\v{s} shows \cite{Tomas:1995} how the homogeneous solution of (\ref{eq_basic_Gamma}) in layer $(j)$ is found in the $(\vkl,\omega)$ Fourier domain (now treating $z$ and $z'$ as parameters and $\vkl$ and $\omega$ as variables, not the other way around)\footnote{For the sake of comparison, ref.\ \cite{Tomas:1995} makes use of the quantity $G_{ik} = \frac{4\pi c^2}{\omega^2}\Gamma_{ik}$}. We have repeated his procedure allowing $\mu(\mb{r})\neq 1$, yielding
\begin{eqnarray*}
  \fl\dyad{\Gamma}_j^{\rmh}(\vkl,\omega;z,z') &=& 
  \frac{\mu_j}{2\kappa_j}\frac{\omega^2}{c^2}\sEM  \frac{\rme^{-\kappa_ja_j}}{D_{qj}}\xi_q\left[ \uv{e}^+_{qj}(\vkl) \rme^{-\kj z}r_{qj}^- \mb{\mathcal{E}}^>_{qj}(-\vkl,\omega;z')\right.\\
  &&+ \left.\uv{e}^-_{qj}(\vkl)\rme^{\kappa_j z}r^+_{qj}\rme^{-\kj a_j}\mb{\mathcal{E}}^<_{qj}(-\vkl,\omega;z')\right].
\end{eqnarray*}
Here $z,z'\in (j)$ and we use the notation $\mb{AB}\equiv \mb{A}\otimes\mb{B}$. The vectors
\begin{eqnarray}
  \mb{\mathcal{E}}^>_{qj}(\vkl,\omega; z) &\equiv \uv{e}^+_{qj}(\vkl)\rme^{-\kappa_j(z-a_j)} + r_{qj}^+ \uv{e}^-_{qj}(\vkl)\rme^{\kappa_j(z-a_j)},\\
  \mb{\mathcal{E}}^<_{qj}(\vkl,\omega; z) &\equiv \uv{e}^-_{qj}(\vkl)\rme^{\kappa_j z} + r_{qj}^- \uv{e}^+_{qj}(\vkl)\rme^{-\kappa_j z}  
\end{eqnarray}
describe waves propagating in layer $(j)$ towards the right and left respectively and which are reflected off the bordering interfaces. Here
\begin{eqnarray*}
  \uv{e}^\pm_{\mbox{TM},j}(\vkl) &= \frac{1}{k_j}(\kl \uv{z} \mp \rmi\kappa_j \uv{k}_\perp) = \uv{e}^\mp_{\mbox{TM},j}(-\vkl)\\
  \uv{e}^\pm_{\mbox{TE},j}(\vkl) &= \uv{k}_\perp \times \uv{z} = -\uv{e}^\mp_{\mbox{TE},j}(-\vkl)
\end{eqnarray*}
are direction vectors in a cartesian coordinate system $(\uv{k}_\perp,\uv{z},\uv{k}_\perp\times\uv{z})$ defined relative to the wave vector:  $\mb{k} = \kl\uv{k}_\perp + k_\parallel \uv{z}$. Futhermore one has introduced the quantities
\begin{eqnarray*}
  \kappa_j &=& \sqrt{\kl^2 - \epsilon_j\mu_j\omega^2/c^2}, \\
  k_j &=& \sqrt{\epsilon_j\mu_j\omega^2/c^2}, \\
  D_{qj} &=& 1-r^+_{qj}r^-_{qj}\rme^{-2\kappa_j a_j}, \\
  \xi_q &=& \delta_{q,\mathrm{TM}}-\delta_{q,\mathrm{TE}}.
\end{eqnarray*}
For physical reasons, recognising that $\left.k_\parallel(z)\right|_{z\in (j)} = \rmi\kappa_j$ we must choose $\im{\kj}<0$. The polarisation mode $q$ runs over the polarisations TE and TM and the Fresnel coefficients $r^\pm_{qj}$ are the relative reflected amplitudes of a $q$-polarised field from the entire stack of layers to the left $(-)$ or right $(+)$ of layer $(j)$.

We may thus write the attraction between multilayers on either side of $(j)$ as (it is sufficient to regard an interface on one side of $(j)$ since attraction is resiprocal)
\be 
  \fl\langle \mathcal{F}_z^0\rangle = \hbar \int_0^\infty \frac{\dd \zeta}{2\pi} \int \frac{\dd^2 \kl}{(2\pi)^2} \left[\epsilon(\Gamma_{xx,j}^{E,\rmh-}+\Gamma_{yy,j}^{E,\rmh-}-\Gamma_{zz,j}^{E,\rmh-})+\frac{1}{\mu}(\Gamma_{xx,j}^{H,\rmh-}+\Gamma_{yy,j}^{H,\rmh-}-\Gamma_{zz,j}^{H,\rmh-}) \right]
\ee
where the Green's function components are taken in the limit $z\to z'\in(j)$ close to either of the interfaces bounding on $(j)$. We assume in this expression that $a_j$ may be varied whereas the thicknesses of other layers are treated as parameters.

The rather complicated expression for $\Gamma^{\mathrm{h}}_{ik,j}$ above may be vastly simplified for our purposes. We introduce ordinary co-ordinates according to the convention of Schwinger et.al.\ \cite{Schwinger:1978} by choosing $\uv{x}=\uv{k}_\perp$ so that $(\uv{k}_\perp,\uv{z},\uv{k}_\perp\times\uv{z}) \to (\uv{x}, \uv{z}, -\uv{y})$. Introducing the important quantity
\be\label{eq_d}
  \frac{1}{d_{qj}} = \frac{r_{qj}^-r_{qj}^+\rme^{-2\kj a_j}}{1-r_{qj}^-r_{qj}^+\rme^{-2\kj a_j}},
\ee
and keeping only terms dependent on $z-z'$ we show with some lengthy but straightforward manipulation that the Green's function may be written very elegantly as
\begin{eqnarray}
  \fl\dyad{\Gamma}^{h-}_j(\vkl,\omega; z,z') = \left[\frac{\kl^2}{\epsilon_j\kappa_j}\frac{1}{d_{\mathrm{TM},j}}\uv{z}\uv{z} -\frac{\kj}{\epsilon_j}\frac{1}{d_{\mathrm{TM},j}}\uv{x}\uv{x} + \frac{\mu_j \omega^2}{\kj c^2}\frac{1}{d_{\mathrm{TE},j}}\uv{y}\uv{y}\right]\cosh(z-z')\nonumber\\
  + \frac{i\kl}{\epsilon_j}\frac{1}{d_{\mathrm{TM},j}}(\uv{z}\uv{x}+\uv{x}\uv{z})\sinh(z-z').\label{eq_Gamma}
\end{eqnarray}

It is now simple matter to calculate $\Gamma^E_{ik}$ and $\Gamma^H_{ik}$  and take the limit $z\to z'$ to find that our final result at zero temperature becomes beautifully simple:
\be
  \langle \mathcal{F}^0(a_j)\rangle = -\frac{\hbar}{2\pi^2}\intz \int_0^\infty \dd\kl \cdot \kl \kj \sEM \frac{1}{d_{qj}}\label{eq_F}.
\ee
At finite temperatures, the integral over all positive imaginary frequencies in (\ref{eq_F}) becomes the sum of the residues of the $\coth$-factors of (\ref{eq_EE}) and (\ref{eq_HH}), the Matsubara frequencies $\rmi\zeta_m = 2\pi \rmi k_BTm/\hbar$,
\be
  \langle \mathcal{F}^T(a_j)\rangle = -\frac{k_B T}{\pi}{\sum_{m=0}^\infty}' \int_0^\infty \dd\kl \cdot \kl \kj \sEM \frac{1}{d_{qj}}\label{eq_FT},
\ee
where the prime on the summation indicates that the $m=0$ term is given half weight.

\section{Generalised reflection coefficients}\label{ch_coefficients}

The task remaining is to evaluate the generalised Fresnel reflection coefficients $r^{\pm}_q$ of a stack of magnetodielectric layers. 

For a single interface between media $(i)$ and $(j)$, the reflected amplitude ratio of a wave arriving from $(i)$ and is partly reflected back into $(i)$ is found from Maxwell's equations to be \cite{Ford:1984}
\be\label{eq_r_ij}
  r_{q,ij} \equiv \Delta_{q,ij} = \frac{\kappa_i - \gamma_{q,ij}\kappa_j}{\kappa_i + \gamma_{q,ij}\kj}
\ee
with $\kappa$ as defined above and 
\[
  \gamma_{q,ij} = \left\{ \begin{array}{cc} \mu_i/\mu_j; &q=\mathrm{TE}\\ \epsilon_i / \epsilon_j; &q=\mathrm{TM}\end{array}\right..
\]

\begin{figure}
  \begin{center}
  \includegraphics[width=3in]{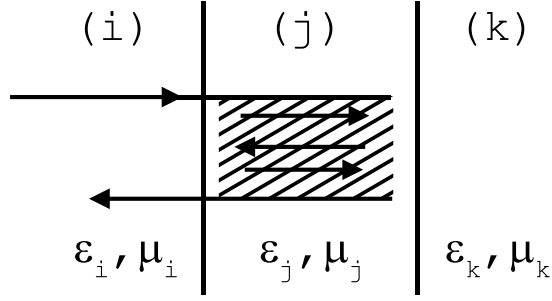}
  \caption{Multiple reflections between interfaces. The scrambled area represents all possible paths originating and ending in layer $(i)$.}\label{fig_ijk}
  \end{center}
\end{figure}

Secondly, the reflection coefficient of a system of two interfaces such as depicted in figure \ref{fig_ijk} may be calculated as the sum of coefficients pertaining to each of the infinitely many optical paths originating and ending in $(i)$. We let the transmission coefficient of a wave transmitted from $(i)$ to $(j)$ be $t_{ij}$ (omitting for now polarisation $q$) and recognise the longitudinal wave vector in $(j)$ to be $k_{\parallel,j} = \sqrt{\epsilon_j\mu_j\omega^2/c^2 - \kl^2} = \rmi\kappa_j$ so that a wave travelling a distance $a_j$ must be multiplied by $\exp(\rmi k_{\parallel,j}a_j) = \exp(-\kj a_j)$; a phase shift if $\kappa_j$ is imaginary (propagating wave) or attenuation if $\kappa_j$ is real (evanescent wave). Thus:
\begin{eqnarray}
  \fl r_{ijk} &= r_{ij} + t_{ij}\rme^{-\kj a_j}r_{jk}\rme^{-\kj a_j}t_{ji} + ... = r_{ij} + t_{ij}t_{ji}r_{jk}\rme^{-2\kj a_j}\sum_{n=0}^\infty (r_{jk}r_{ji}\rme^{-2\kj a_j})^n\nonumber \\
  \fl&=\frac{r_{ij} + r_{jk}\rme^{-2\kj a_j}}{1+r_{ij}r_{jk}\rme^{-2\kj a_j}}\label{eq_r_ijk}
\end{eqnarray}
where we have made use of the properties $t_{ij}t_{ji}-r_{ij}r_{ji} =1$ and $r_{ij}=-r_{ji}$. Equation (\ref{eq_r_ijk}) is valid for either polarisation respectively.

This provides a simple procedure for calculating the reflection coefficient of a multilayer containing any finite number of interfaces. To calculate $r^-_{qj}$, say, as it appears in figure \ref{fig_multilayer}, we start with the leftmost interface between zones $(1)$ and $(0)$ and find $r_{q,10}=\Delta_{q,10}$ and invoke (\ref{eq_r_ijk}) recursively to find the reflection coefficient of the two leftmost interfaces, then the three leftmost and so on until the closest interface, between $(j)$ and $(j-1)$, is reached. 

\section{The Casimir attraction calculated in various configurations}\label{ch_configs}

We go on to demonstrate the strength of the above procedure by calculating the Casimir force in an array of different plane parallel configurations so as to reproduce the results of several references.

\subsection{Two half-spaces}

\begin{figure}
  \begin{center}
  \includegraphics[width=3.5in]{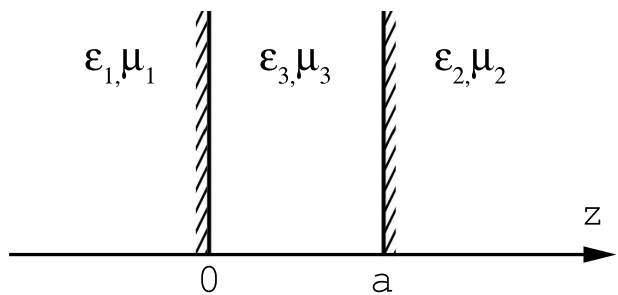}
  \caption{The trizone configuration of two half-spaces separated by a gap region.}\label{fig_trizone}
  \end{center}
\end{figure}

Consider first the simplest system of two half-spaces of some magnetodielectric material separated by a gap of width $a$, generally made of some other material. We denote the half-spaces 1 and 2 (see figure \ref{fig_trizone}) and the gap 3.

There is now only one interface on either side of the gap and obtaining the force expression is almost trivial. The reflection coefficients to the right and left are
\[
  r^+_q = \Delta_{q,32} \hspace{20pt}\mathrm{ and }\hspace{20pt} r^-_q = \Delta_{q,31}
\]
as defined in (\ref{eq_r_ij}). We get the force density expression (suppressing the averaging notation henceforth)
\be
  \mathcal{F}^0(a) = -\frac{\hbar}{2\pi^2}\intz \int_0^\infty \dd\kl \cdot \kl \kappa_3 \sEM \frac{\Delta_{q,32}\Delta_{q,31}\rme^{-2\kappa_3 a}}{1-\Delta_{q,32}\Delta_{q,31}\rme^{-2\kappa_3 a}},
\ee
which is the classical Lifshitz result quoted in numerous references, e.g.\ \cite{Lifshitz:1961,Schwinger:1978}.

\subsection{A plate outside a wall}

\begin{figure}
  \begin{center}
  \includegraphics[width=2.3in]{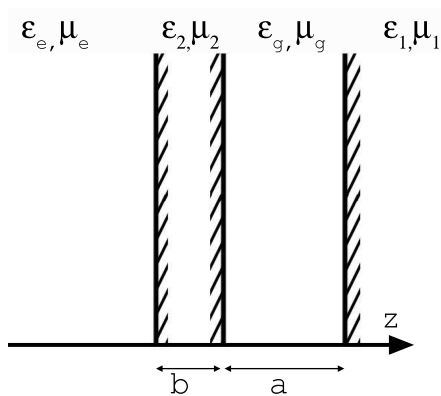}
  \caption{The four-layered configuration of a plate and a wall.}\label{fig_4z}
  \end{center}
\end{figure}

We consider a plate of finite thickness $b$ and material denoted $2$ separated by a distance $a$ from an infinitely thick wall of material $1$ as depicted in figure \ref{fig_4z}. Such a system has recently been considered by Toma\v{s} \cite{Tomas:4z} for the sake of discussing the consequences of using an alternative Lorentz force stress tensor, and his results using the Minkowski tensor agrees with ours. The gap material, denoted with subscript $g$ is allowed to be different from the material of the exterior, denoted with subscript $e$. The Green's function is calculated in the gap region, and $r_q^-$ and $r_q^+$ are the relative reflected amplitude of a wave originating in the gap and propagating towards the left and right respectively.

We see immediately that $r^+_q=(\kappa_g-\gamma_{q,g1}\kappa_1)/(\kappa_g+\gamma_{q,g1}\kappa_1)=-\Delta_{q,1g}$  and combining (\ref{eq_r_ij}) and (\ref{eq_r_ijk}) readily yields
\[
  r_q^- = \frac{-\Delta_{q,2g}+\Delta_{q,2e}\rme^{-2\kappa_2 b}}{1-\Delta_{q,2g}\Delta_{q,2e}\rme^{-2\kappa_2 b}},
\]
so that
\[
   \frac{1}{d_{qj}} = \frac{\left(\Delta_{q,1g}\Delta_{q,2g}-\Delta_{q,1g}\Delta_{q,2e}\rme^{-2\kii b}\right)\rme^{-2\kg a}}{1-\Delta_{q,1g}\Delta_{q,2g}\rme^{-2\kii b} -\left(\Delta_{q,1g}\Delta_{q,2g}-\Delta_{q,1g}\Delta_{q,2e}\rme^{-2\kii b}\right)\rme^{-2\kg a}},
\]
from which the Green's function and Casimir attraction per unit area between plate and wall follow neatly from (\ref{eq_Gamma}) and (\ref{eq_F}).

\subsection{A slab in a cavity}

\begin{figure}
  \begin{center}
    \includegraphics[width=3in]{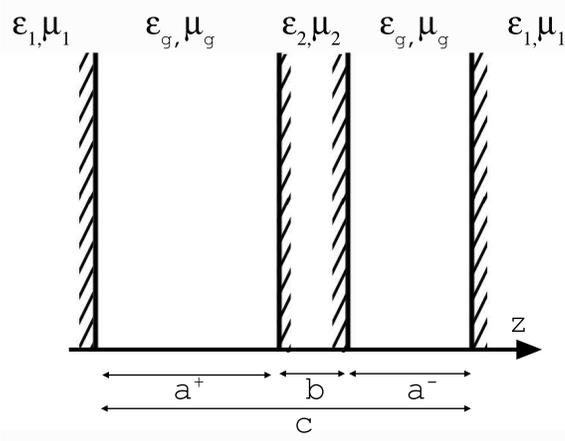}
    \caption{The system of a slab in a cavity.}\label{fig_5z}
  \end{center}
\end{figure}

We go on to study the five-layered system of a slab between two walls as defined in figure \ref{fig_5z}. A general system of five layers was first considered by Zhou and Spruch \cite{Zhou:1995} and the special case of a slab in a cavity was treated, apparently using the method presented here, by Toma\v{s} \cite{Tomas:2002}. Some manipulation shows that our result co-incide with the finds of both references. We calculate the attraction with respect to each of the gaps in turn and subsequently find the force acting on the slab as the difference between these. The procedure is identical to that above so the details of the somewhat more lengthy but principally uncomplicated calculations are left out.

We denote the left and right hand gaps with superscript $+$ and $-$ respectively for reasons which will become obvious and for simplicity we assume both walls to be made of the same material, denoted $1$, whereas the slab is made of a material indexed $2$. Furthermore, we use the simplifying notation $\Delta_{iq} \equiv \Delta_{q,ig}$ with $q=1,2$. With this we find for the left and right hand gaps 
\[
  \frac{1}{d_q^\pm} = \frac{U_q^\mp \rme^{-2\kg a^\pm}}{V_q^\mp-U_q^\mp \rme^{-2\kg a^\pm}}
\]
where
\begin{eqnarray*}
  U_q^\pm &=& \Diq\Diiq (1-\Diq\Diiq \rme^{-2\kg a^\pm}) - \Diq(\Diiq - \Diq \rme^{-2\kg a^\pm})\rme^{-2\kii b},\\
  V_q^\pm &=& 1-\Diq\Diiq \rme^{-2\kg a^\pm} - \Diiq(\Diiq - \Diq \rme^{-2\kg a^\pm})\rme^{-2\kii b},
\end{eqnarray*}
and resulting force on the centre slab is accordingly (a function of \emph{either} $a^+$ or $a^-$ when $b$ and $c$ are assumed constant parameters)
\be
  \mathcal{F}^0(a^\pm; b,c) = \frac{\hbar}{2\pi^2}\int_0^\infty \dd \zeta \int_0^\infty \dd \kl\cdot \kl\kg \sEM \left( \frac{1}{d_q^-}-\frac{1}{d_q^+}\right).
\ee
Notice how, if we let the gap be either very wide ($b\to\infty$) or perfectly reflecting ($\kappa_2 \to \infty$), the terms contaning the factor $\exp(-2\kappa_2 b)$ vanish, and we get back Lifshitz' expression for two separate gaps as we should.

A rather more instructive expression is obtained if the position of the slab is given not by $a^+$ and $a^-$, but the deviation $\delta$ of the centre of the slab from the midline of the cavity. Introducing the quantity $h = c-b = a^++a^-$, we write $a^\pm = h/2\pm \delta$ and the force density expression becomes after some shuffling of symbols
\be
  \mathcal{F}^0(\delta; b,c) = \frac{\hbar}{2\pi^2}\intz \int_0^\infty \dd\kl\cdot \kl\kg \sEM \frac{A_q\sinh 2\kg \delta}{B_q-A_q \cosh 2\kg\delta}
\ee
with
\begin{eqnarray*}
  A_q &=& 2\Diq\Diiq(1-\rme^{-2\kii b})\rme^{-\kg h},\\
  B_q &=& 1-\Diiq^2\rme^{-2\kii b} + \Diq^2 (\Diiq^2 - \rme^{-2\kii b})\rme^{-2\kg h}.
\end{eqnarray*}

One should note in this context that this is the geometry treated by Raabe and Welsch \cite{Raabe:2005} and later by Toma\v{s} \cite{Tomas:5z}, making use of their alternative, Lorentz-type stress tensor. 

\subsection{Two plates of finite thickness}

The system of two plates each of finite thickness has been treated by numerous authors. First to do so was Kupiszewska \cite{Kupiszewska:1990}, who employed an effectively one-dimensional model by insisting that waves be reflected at normal incidence, as have several authors after her. A three dimensional geometry, allowing nonzero values of $\kl$, appears first to have been considered by Jaekel and Reynaud in 1991 \cite{Jaekel:1991} whose result is found to be smaller than ours by a factor $1/2$ (it should be noted that comparison is not trivial due to formal differences). A number of other references \cite{Zhou:1995, Matloob:2001, Mochan:2002}, however, obtain results agreeing perfectly with that found by using the above procedure. The principally identical system of two semispaces each covered with a thin layer of a different substance was considered by Klimchitskaya and co-workers \cite{Klimchitskaya:2000}, again in agreement with the below.

\begin{figure}
  \begin{center}
    \includegraphics[width=3.5in]{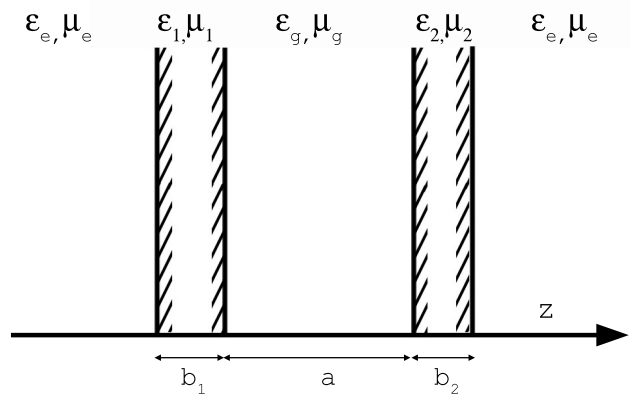}
    \caption{A system of two plates of finite thickness.}\label{fig_2p}
  \end{center}
\end{figure}

The geometry is defined in figure \ref{fig_2p}. Just as before we determine the reflection coefficients of either slab as seen from the gap,
\begin{equation}
  r^+_q = \frac{-\Delta_{q,2g}+\Delta_{q,2e}\rme^{-2\kii b_2}}{1-\Delta_{q,2g}\Delta_{q,2e}\rme^{-2\kii b_2}}\hspace{30pt}
  r^-_q = \frac{-\Delta_{q,1g}+\Delta_{q,1e}\rme^{-2\ki b_1}}{1-\Delta_{q,1g}\Delta_{q,1e}\rme^{-2\ki b_1}},
\end{equation}
and the Casimir attractive force per unit area of the plates is found neatly from (\ref{eq_d}) and (\ref{eq_F}).

\section{The neglected bulk force}

In the case that materials $0$ and $n$ in figure \ref{fig_multilayer} are different, a second force appears in addition to the Casimir force. It is a force acting on the entire multilayer system and we shall refer to it as a bulk force. Quantitatively the electromagnetic force density was found to be $\mb{f} = \dyad{T}\cdot \buildrel{\leftarrow}\over{\nabla}$ and the force acting on some volume $\mathcal{V}$ is 
\[
  \mb{F} = \int_\mathcal{V} \dd^3 r \mb{f} = \oint_{\partial \mathcal{V}} \dyad{T}\cdot \dd \mb{S}
\]
where the divergence theorem has been invoked and $\dd \mb{S}$ points normally out of $\mathcal{V}$, enclosed by the surface $\partial\mathcal{V}$. Let $\mathcal{V}$ be a box enclosing all interfaces of the multilayer system so that its $z$-boundaries lie at $z^-\in (0)$ and $z^+\in(n)$. Only the sides of $\mathcal{V}$ parallel to the $xy$-plane contribute to the bulk force, which is evaulated per unit transverse area as (averaging with respect to fluctuations is understood)
\begin{eqnarray*}
  \fl\bf{\mathcal{F}}_\mathrm{bulk} &=& T_{zz}(z^+)-T_{zz}(z^-)\cdot \uv{z}\\
  \fl&=& -\frac{\uv{z}\epsilon_0}{2}\left[\epsilon(z^+)\left.(E_z^2-E_x^2-E_y^2)\right|_{z=z^+}-\epsilon(z^-)\left.(E_z^2-E_x^2-E_y^2)\right|_{z=z^-} \right]\\
  \fl&&-\frac{\uv{z}\mu_0}{2}\left[\mu(z^+)\left.(H_z^2-H_x^2-H_y^2)\right|_{z=z^+}-\mu(z^-)\left.(H_z^2-H_x^2-H_y^2)\right|_{z=z^-} \right].
\end{eqnarray*}
If now media $(0)$ and $(n)$ are the same, the mean squared fluctuating fields will be the same on either side of the multilayer, and the bulk force is zero, otherwise $\mathcal{F}_\mathrm{bulk}$ is generally nonzero.

This force is typically neglected, and an argument in favour of doing so is surely that in a real system, layers are not infinitely thick. The outmost layers of a multislab configuration should realistically be air or vacuum, and if they are not, it simply means that $\mathcal{V}$ does not contain the entire system and the bulk force is identically cancelled by reflections at surfaces (not necessarily parallel with the system, or even plane) outside $\mathcal{V}$, as it should be according to Newton's third law.

\section{Conclusion}

With the above procedure, the calculation of Casimir forces in even complex multilayered geometries is both quick and straightforward and shown able to reproduce the results of a number of previous works. Various configurations may thus be considered theoretically and numerically with ease to study the various dependencies e.g.\ on material properties and temperatures. The procedure may furthermore be repeated to reveal differences between various electromagnetic stress tensors, a subject of dispute for decades \cite{Brevik:1979, Stallinga:2006}.

\section*{Acknowledgements}

The author thanks professor Iver Brevik for helpful comments and many rewarding discussions on the subject.

\end{document}